\documentclass[journal,onecolumn,12pt]{IEEEtran}
\usepackage{array}
\usepackage{ulem}

\ifCLASSINFOpdf
\else
\fi
\usepackage{supertabular}
\usepackage{amsfonts}
\usepackage{threeparttable}

\begin{document}
%
\title{On Disjoint Golomb Rulers}
%
%
%

\author{Baoxin Xiu,
        Changjun Fan,
        and Meilian Liang 
\thanks{Baoxin Xiu and Changjun Fan are with the School of Information System and Management, National University of Defense Technology, Changsha 410073, P.R.China .}
\thanks{Meilian Liang is with the School of Mathematics and Information Science, Guangxi University, Nanning 530004, P.R.China. (Corresponding author. E-mail: ElaineGXU@gmail.com)}
\thanks{This work is Partially supported by the NSF of China (11361008).}}

%
%

\markboth{}%
{Xiu \MakeLowercase{\textit{et al.}}: On Disjoint Golomb Ruluers}
%



\maketitle

\begin{abstract}
A set $\{a_i\:|\: 1\leq i \leq k\}$ of non-negative integers is a Golomb ruler if differences $a_i-a_j$, for any $i \neq j$, are all distinct. A set of  $I$ disjoint Golomb rulers (DGR) each being a $J$-subset of $\{1,2,\cdots, n\}$ is called an $(I,J,n)-DGR$.
Let $H(I, J)$ be the least positive $n$ such that there is an $(I,J,n)-DGR$. In this paper, we propose a series of conjectures on the constructions and structures of DGR. The main conjecture states that if $A$ is any set of positive integers such that $|A| = H(I, J)$, then there are $I$ disjoint Golomb rulers, each being a $J$-subset of $A$, which generalizes the conjecture proposed by Koml{\'o}s, Sulyok and Szemer{\'e}di in 1975 on the special case $I = 1$. These conjectures are computationally verified for some values of $I$ and $J$ through modest computation. Eighteen exact values of $H(I,J)$ and ten upper bounds on $H(I,J)$ are obtained by computer search for $7 \leq I \leq 13$ and $10 \leq J \leq 13$. Moveover for $I > 13$ and $10 \leq J \leq 13$, $H(I,J)=IJ$ are determined without difficulty.
\end{abstract}

\begin{IEEEkeywords}
Golomb ruler, Sidon set, disjoint Golomb rulers.
\end{IEEEkeywords}

%
\IEEEpeerreviewmaketitle

\newtheorem{df}{Definition}
\newtheorem{thm}{Theorem}
\newtheorem{clm}{Claim}
\newtheorem{rem}{Remark}
\newtheorem{lem}{Lemma}
\newtheorem{cor}{Corollary}
\newtheorem{conj}{Conjecture}

\section{Introduction}\label{intro}
%
%
%
%
\IEEEPARstart{A}  {} $k$-mark  Golomb ruler is a set of $k$ distinct non-negative integers, also called marks, $\{a_i\:|\: 1\leq i \leq k\}$ such that all differences $a_i-a_j, i \neq j$ are distinct. The difference between the maximal and minimal integer is referred to as the length of the Golomb ruler. Golomb rulers give various important applications in engineering, for example, the radio-frequency allocation for avoiding third-order interference\cite{Babcock, Golomb1977}, the construction of convolutional or LDPC codes\cite{Robinson2, Chen2012}, the design of recovery schemes for faulty computers\cite{klonowska2005optimal}, etc.

With wide applications in the real world and an inherent mathematical interest, Golomb rulers have been extensively studied by mathematicians and computer scientists. Various algebraic methods have been proposed to construct Golomb rulers as densely populated with marks as possible\cite{ET1941,ruzsa1993solving,bose1962theorems,singer1938theorem,drakakis2009symmetry}. However, \textit{Optimal Golomb Rulers}, each of which is the shortest Golomb ruler possible for a given number of marks, can only be discovered or verified by exhaustive computer search.  For the highly combinatorial nature of Golomb rulers, lengths of $k$-mark optimal rulers have been determined only for $k \leq 27$ so far\cite{WikiUrL}. The Golomb ruler problem has been used as a standard benchmark for Artificial Intelligence research\cite{meyer2009complexity}. Dual to Golomb rulers in a certain way, \textit{Sidon Sets} which are subsets of $\{1,2,\cdots,n\}$ with distinct pairwise sums of elements\cite{Sidon}, have been studied independently in combinatorial number theory for decades, producing a large amount of results, among which a few seem little known to the Golomb ruler community.

The problem of finding disjoint Golomb rulers (abbreviated as DGRP), a generalization of the Golomb ruler problem, was first considered by Chen\cite{wende1981} in mobile radio-frequency allocation for a collection of areas avoiding third-order interference within each area. We say that a set of  $I$ disjoint Golomb rulers (abbreviated as DGR) each being a $J$-subset of $\{1,2,\cdots, n\}$ is an $(I,J,n)$-DGR. $H(I, J)$ is defined to be the least positive integer $n$ such that there is an $(I,J,n)$-DGR. To determine $H(I,J)$ for all $I$ and $J$ is an extremely challenging task. Kl{\o}ve\cite{Klove} proposed a number of constructions for DGR and gave a table of exact values and bounds on $H(I,J)$ for $I \leq 11, J \leq 9$, which was improved and extended through computer search by Shearer \cite{Shearer}.

In \cite{Shearer}, DGRP was transformed into a problem of finding independent sets in a graph with $J$-mark Golomb rulers  from $\{1, 2, \cdots, n\}$  as vertices and edges between pairs of joint rulers, and thus had been solved perfectly for small $I$ and $J$. But due to the huge amount of such Golomb rulers for large $J$ and $n$ (for example, there are already more than 8 million 10-mark Golomb rulers chosen from $\{1, 2, \cdots, 80\}$) , constructing a graph with all rulers as vertices may lead to high or even unaffordable storage and time consumption. Besides, the independent determination of $H(I,J)$ will leave out some useful information hidden in $(I',J,H(I',J))$-DGR $(I' < I)$, which may speed up the search process. For large $I$ and $J$, an imitation of the method without much innovation is inadequate to determine $H(I,J)$, and to determine lower bounds on $H(I,J)$ by complete search is extremely difficult. For a partial search to determine upper bounds on $H(I,J)$, it is quite critical to specify the starting vertices and the traversing order of the vertices, and to restrict the search radius and time, which depend on our knowledge or speculation about the existence and the distribution of DGR.

Among DGR, we are, in particular, interested in those $(I,J,n)$-DGRs such that $n=H(I,J)=IJ$, which we call regular. In \cite{Klove}, Kl{\o}ve proved that there exists a bound $\iota(J)$ such that regular DGR exist for all $I \geq \iota(J)$. He also gave values of $\iota(J)$ for $J \leq 6$ and bounds on $\iota(J)$ for $7 \leq J \leq 16$, among which exact values of $\iota(J)(J=7,8,9)$ later were determined by Shearer \cite{Shearer}. Although new constructions for regular DGR were proposed in  \cite{chen1998new}, bounds on $\iota(J)$ have not been substantially improved.

This paper proposes a series of conjectures on DGR, based on which eighteen exact values of $H(I,J)$ and ten upper bounds on $H(I,J)$ are obtained, where $7 \leq I \leq 13, 10 \leq J \leq 13$. Some exact values of $H(I,J)$ for $I > J$ are also determined, narrowing the bounds on $\tau(J)$. Deduced from a conjecture on DGR, a conjecture on optimal Golomb rulers is given as an incidental finding.

The remainder of the paper is organized as follows. Section \ref{secmaincon} proposes a series of conjectures on DGR. Section \ref{secmoreconj} gives one more conjecture on DGR, which deduces a conjecture on  optimal Golomb rulers. Section \ref{computation} proposes a novel method of determining upper bounds on $H(I,J)$ and presents computational results of verification of these conjectures. We conclude with a summary in Section \ref{conclu}.

\section{Some Conjectures on Disjoint Golomb Rulers}\label{secmaincon}

We propose some conjectures on disjoint Golomb rulers in this section.

\subsection{The Main Conjecture}

The Golomb ruler problem can be generalized to arbitrary $n$ integers (not necessarily the first $n$). Let $A = \{a_1, \cdots, a_n\}$ be an arbitrary  finite set of positive integers. Koml{\'o}s, Sulyok and  Szemer{\'e}di  argued in \cite{kssa1975} that $A$ may contain an $m$-mark Golomb ruler with $m = (1+o(1))n^{1/2}$. They further proved that $m > cn^{1/2}$ for a certain positive constant $c$. The constant was improved in \cite{RuzsaII1995}.

As pointed out in \cite{kssa1975}, one can expect that the case of the first $n$ positive integers is ``the worst case", which can be generalized to the following conjecture.

\vspace{0.3cm}
\begin{conj} \label{genefirstn}
Suppose that $I$ and $J$ are positive integers. If $A$ is any set of positive integers such that $|A| = H(I, J)$, then there are $I$ disjoint Golomb rulers, each being a $J$-subset of $A$.
\end{conj}
\vspace{0.3cm}

It is not difficult to see that if $I=1$, then Conjecture \ref{genefirstn} coincides with the idea in \cite{kssa1975}.

\subsection{More Conjectures on Disjoint Golomb Rulers}

If we can prove Conjecture \ref{genefirstn}, then we can prove more interesting results based on it.  The idea of the following conjecture will be used in computing upper bounds on $H(I,J)$  for small $I$ and $J$ in this paper.

\vspace{0.3cm}
\begin{conj} \label{difference}
If $I$ and $J$ are integers, and $I \geq 1$, $J \geq 3$, then $H(I+1, J) \leq H(I, J) + J$.
\end{conj}
\vspace{0.3cm}

It seems interesting to show whether the inequality $H(I+1, J) \leq H(I, J) + J$ in Conjecture \ref{difference} strictly holds when $H(I, J) > IJ$.

Among DGR, we are very interested in regular ones. For $I_0 \geq 2$, we may propose the following Conjecture \ref{genefirstn2}.

\vspace{0.3cm}
\begin{conj} \label{genefirstn2}
If $H(I_0, J) = I_0 J$, then $H(I, J) = IJ$ for any integer $I > I_0$.
\end{conj}
\vspace{0.3cm}

If Conjecture \ref{genefirstn} holds, then the following conjecture holds too.

\vspace{0.3cm}
\begin{conj} \label{genefirstnb}
For any Golomb ruler $A_1 \subseteq \{1, 2, \cdots, (I+1)J\}$ such that $|A_1| = J$, if $H(I,J) = IJ$ then there exists a  regular $(I+1,J,(I+1)J)$-DGR containing $A_1$.
\end{conj}
\vspace{0.3cm}

We can see that if Conjecture \ref{genefirstnb} holds then Conjecture \ref{genefirstn2} holds too.

\subsection{Theorems on Conjectures}

In the following two theorems, we will prove some results on conjectures previously proposed.
\vspace{0.3cm}
\begin{thm}\label{genefirstn2to3}
If Conjecture \ref{difference} holds, then Conjecture \ref{genefirstn2} holds.
\end{thm}
\vspace{0.3cm}
\begin{IEEEproof}
If Conjecture \ref{difference} holds and $H(I_0, J) = I_0 J$, then $(I_0+1)J \leq H(I_0+1, J) \leq H(I_0,J)+J = I_0J+ J = (I_0+1)J$. So $H(I_0+1, J) = (I_0+1)J$. Thus by Mathematical Induction we can see that if Conjecture \ref{difference} holds, then Conjecture \ref{genefirstn2} holds too.
\end{IEEEproof}
\vspace{0.3cm}

\begin{thm}\label{genefirstn1to2}
If Conjecture \ref{genefirstn} holds, then Conjecture \ref{difference} holds.
\end{thm}
\vspace{0.3cm}
\begin{IEEEproof}
Suppose that $H(I, J) = n$  and  $A = \{1, 2, \cdots, n+J\}$.  Let $A_0$ be any subset of $A$ such that $A_0$ is a Golomb ruler and $|A_0| =J$. So $|A - A_0| = n$.
If Conjecture \ref{genefirstn} holds, then $A - A_0$ is the union of $I$ disjoint $J$-mark Golomb rulers. Suppose that these Golomb rulers are $\{A_j \;|\; 1 \leq j \leq I\}$.
So  $\{A_j \;|\; 0 \leq j \leq I\}$ are $I$ disjoint $J$-mark Golomb rulers, all of which are subsets of $A$. Thus $H(I+1, J) \leq H(I, J) + J$. Thus Conjecture \ref{difference} holds.
\end{IEEEproof}
\vspace{0.3cm}

Based on Theorem \ref{genefirstn2to3} and Theorem \ref{genefirstn1to2}, it is not difficult to see that the following theorem holds.
\vspace{0.3cm}
\begin{thm}\label{genefirstn1to3}
If Conjecture \ref{genefirstn} holds, then Conjecture \ref{genefirstn2} holds.
\end{thm}

\vspace{0.3cm}

Imitating the definition of $H(I,J)$, we define $Y(I,J)$ to be the smallest $n$ such that any set of positive integers with $n$ elements, contains $I$ disjoint $J$-mark Golomb rulers.

By the results in \cite{kssa1975} we know that $Y(1,J)$ exists for any positive integer $J$, and we can obtain a better upper bound on $Y(1,J)$ by the results in \cite{RuzsaII1995}. Similar to the proof of Theorem \ref{genefirstn1to2}, we can prove $Y(I+1, J) \leq Y(I, J) + J$ for any integers $I \geq 1$ and $J \geq 3$.

\section{More Conjectures on Disjoint Golomb Rulers and Optimal Golomb Rulers} \label{secmoreconj}

Let $G(k)$ be the length of an optimal Golomb ruler with $k$ marks. Singer proved that if $q$ is a power of a prime, then there exist $q + 1$ integers that have distinct differences modulo $q^2 + q + 1$ and thus
form a Golomb ruler\cite{singer1938theorem}, which implies that $G(n) < n^2-n+1$ if $n-1$ is a prime power.

Based on the Singer construction, the following results were proved in \cite{Klove}.
\vspace{0.3cm}
\begin{thm}
If $p$ is a prime power, then
$H(p+1, p) = p^2+p$, $H(p, p-1) \leq p^2-2$, $H(p-1, p) \leq p^2-1$.
\end{thm}
\vspace{0.3cm}

The first equation in Theorem 4 is valid for regular DGR. We propose the following conjecture on $H(I, I+2)$.

\vspace{0.3cm}
\begin{conj} \label{HII2regular}
For any integer $I \geq 8$, $H(I, I+2) = I(I+2)$.
\end{conj}
\vspace{0.3cm}

If both Conjecture \ref{genefirstn2} and Conjecture \ref{HII2regular} hold, then we have that $H(I, J) = IJ$ for any integer $I \geq J-2$.

Moreover, if both Conjecture \ref{genefirstn} and Conjecture \ref{HII2regular} hold, then for any two disjoint $J$-mark Golomb rulers from $\{1, 2,  \cdots, IJ\}$, $g_1$ and $g_2$, we can always discover a regular $(I,J,IJ)$-DGR where $6 \leq J \leq I$ containing $g_1$ and $g_2$. Note that it may be time consuming to computationally verify this even for small $I$ and $J$.

Upper bounds given by Singer work only for a prime (or power of a prime) number of marks. But if Conjecture \ref{HII2regular} holds, it is not difficult to see that the following bounds on $G(k)$ hold for general situations .
\vspace{0.3cm}
\begin{conj} \label{SRkk+2b}
$G(k+2) < k^2 + k$ for any $k \geq 6$.
\end{conj}
\vspace{0.3cm}

It can be seen that Conjecture \ref{SRkk+2b} is stronger than the old conjecture that $G(k) < k^2$ for all $k > 0$, which was first mentioned by Erd\H{o}s in an equivalent form\cite{ET1941} and is known to be true for $k \leq 65000$ up to now\cite{dimitromanolakis2002analysis}.

Conjecture \ref{SRkk+2b} holds for any $k \in \{8, \cdots, 150\}$, which was confirmed based on Golomb rulers shown in \cite{ShearUrL1}. Additional computational verification of the conjecture is not conducted.

\section{Upper Bounds on $H(I,J)$ and Computational Verification of Conjectures} \label{computation}

\subsection{Basic Notations}

We first introduce some notations that will be used in this section.

Let $D =\{\Delta_1, \Delta_2, \cdots, \Delta_I\}$ be an $(I,J,n)$-DGR. Note that $\Delta_i(i=1,2,\cdots,I)$ does not necessarily contain 1 or $n$.  Let $\sigma(D ) = \min\{a_{ij} \:| \:1 \leq i \leq I, 1 \leq j \leq J, a_{ij} \in \Delta_i, \Delta_i \in D \}$, $\lambda(D ) = \max\{a_{ij}\: |\: 1 \leq i \leq I, 1 \leq j \leq J, a_{ij} \in \Delta_i, \Delta_i \in D \}$.

\vspace{0.3cm}
\begin{df}
Define $D +b = \{\{a_{ij}+b \:| \: 1 \leq j \leq J , a_{ij} \in \Delta_i\} \: | \: \Delta_i \in D, 1 \leq i \leq I \}$  to be the $b$-step transformation of $D$.
\end{df}
\vspace{0.3cm}

Intuitively, $D +b$ corresponds to shifting all Golomb rulers in $D$ right (or left) by $|b|$ when $b \geq 0$ (or  $b < 0$). Observe that $D+b$ is an $(I,J,m)$-DGR for any $b \in \{1-\sigma(D ), \cdots, m-\lambda(D )\}$, where $m \geq n$.

\vspace{0.3cm}
\begin{df}
$\mathcal{S} (D ,m) = \{D +i \ \:|\: 1-\sigma(D ) \leq i \leq  m-\lambda(D )\}$ is the transformation set of $D$ within $m$.
\end{df}
\vspace{0.3cm}

Obviously, any $k$-subset of $D$($1 \leq k \leq I$) , denoted by $A$, is a $(k,J,n)$-DGR. We are more interested in the transformation set of $A$ than that of $D$ .

\vspace{0.3cm}
\begin{df}
 Let $\mathcal{R}_0$ be the full set of all $(I,J,n)$-DGR and $\mathcal{R} \subseteq \mathcal{R}_0$.

 $ \mathcal{T} (\mathcal{R} ,k,m) = \bigcup_{D \in \mathcal{R}}\bigcup_{ A \in P(D ), |A | = k} \mathcal{S} (A ,m)$ is the $k$-sub transformation set of $\mathcal{R}$ within $m$,  where $P(D )$ is the power set of the set $D $.
\end{df}
\vspace{0.3cm}

The set $\mathcal{R}$ is a collection of $(I,J,n)$-DGR, and $\mathcal{T} (\mathcal{R} ,k,m)$ is a collection of $(k,J,m)$-DGR transformed from elements in $\mathcal{R}$.

\subsection{Upper Bounds on H(I,J) for Some Values of $I$ and $J$}

For the convenience of illustration, let $\mathcal{G}^J_n$  be the graph constructed by the method in \cite{Shearer}, with $J$-mark Golomb rulers  from $\{1,2,\cdots,n\}$ as vertices. Observe that for an independent set $D$ of size $I$ in $\mathcal{G}^J_n$,  any $\xi \in \mathcal{T} (\{D\}, k, m)$ is also an independent set in $\mathcal{G}^J_m$, where $m > n$ and $1 \leq k \leq I$.  Starting from $\xi$ and restricting the search on the common non-neighborhood of vertices in $\xi$ may speed up the search for $(I+1,J,m)$-DGR. Furthermore, to obtain the upper bound on $H(I+1,J)$, we may assume the upper bound is $H(I,J)+J$ (or even smaller) firstly,  following the idea behind Conjecture \ref{difference}, and then try to verify it and improve it.

Suppose that $n$ is the best known upper bound on $H(I,J)$ and $\mathcal{R}$ is a set of selected $(I,J,n)$-DGRs. Given an $m>n$, the procedure of finding $(I+1,J,m)$-DGR is as follows. First, a $k \in \{1,2,\cdots,I-1\}$ is considerably selected. Then for every $\xi \in \mathcal{T}(\mathcal{R},k, m)$, all Golomb rulers containing $J$ elements chosen from $\{1,2,\cdots,m\} - \bigcup_{\Delta \in \xi} \Delta$ are listed, on which an exhaustive search is performed to find a collection, denoted by $A$, of $I+1-k$ disjoint rulers. If $A$ is not found, the same search process is repeated on $\mathcal{T}(\mathcal{R},k',m)$ where $k'<k$ until $A$ has been found or time is up. When $A$ is found, $m$ is determined to be the upper bound on $H(I+1,J)$ and $\xi \bigcup A$ is an $(I+1,J,m)$-DGR. Note that finding $I+1-k$ DGR is a hard and time consuming work if $I+1-k$ is large. We did not use the method of constructing graphs in \cite{Shearer}. However, we applied a number of small programming tricks to improve the efficiency of the search process, which are not special to be addressed in detail here.

The aforementioned  procedure of finding DGR has to be repeated until an ideal upper bound on $H(I+1,J)$ is determined.  An $m_0 \in \{a,a+1,\cdots, n+J\}$ is specified first, where $a = \max \{n+1, (I+1)J \}$. If an $(I+1,J,m_0)$-DGR is found, $m \in \{a,a+1,\cdots,m_0-1\}$ is checked one by one in a descending order until $m=a$ or no $(I+1,J,m)$-DGR is found. If  $m=a$, $a$ is the exact value of $H(I+1,J)$; otherwise, $m+1$ is the best upper bound on $H(I+1,J)$ we can obtain for a given $\mathcal{R}$. If no $(I+1,J,m_0)$-DGR is found, values greater than  $m_0$ are tried until an $(I+1,J,m')$-DGR where $m' > m_0$ is found, bringing the upper bound on $H(I+1,J)$. The upper bound can be verified as the exact value only through exhaustive computer search, which is only performed for a few cases in this paper.

Let us illustrate the procedure by the example of computing $H(I,10)$ based on $(I-1, 10, H(I-1,10))$-DGR $(I = 7,8,9,10)$, with key details depicted in Table \ref{table:algo}. For that $H(6,10)=70$, to determine $H(7,10)$ all distinct $(6,10,70)$-DGRs are first obtained by computer search in a very short time, which form the set $\mathcal{R}$. Then by the assumption of $H(8,10)=80$ based on Conjecture \ref{HII2regular}, $m$ is initialized to be $75$, which is the midpoint between 71 and 79. Note that the initial value of $m$ can be chosen somewhat casually from the range. Given $k=3$, a $(7,10,75)$-DGR is quickly discovered. The next search for $(7,10,74)$-DGR does not succeed until $k=1$ (shown in the third column). A Golomb ruler $\Delta = \{12, 13, 18, 22, 35, 38, 46, 53, 65, 67\}$ is chosen (underlined and shown in the second column) and after it is shifted left by two (that is $b=-2$, shown in the fourth column), six DGR from $\{1,2,\cdots,74\}-\Delta$ are finally found. Thus a $(7,10,74)$-DGR is discovered, which is the second block shown in the second column, with the (-2)-step transformation of $\{\Delta\}$ in bold. Then for the failure of the partial search for $(7,10,73)$-DGR, a complete search is performed but none is found, verifying $H(7,10)=74$. Again three Golomb rulers are chosen from the $(7,10,74)$-DGR just found and are shifted right by one. Through a similar process, an $(8,10,80)$-DGR is found, which is regular. So $H(8,10)$ is determined to be $80$. Thereafter, $H(9,10)=90$ and $H(10,10)=100$ are determined easily by the same way .

Note that some values of $H(I,J)$ can even be determined by an $(I-i,J, H(I-i,J))$-DGR where $i \geq 2$. For example, a $(10,11,110)$-DGR is obtained using an 11-step transformation of the set of the last four rulers of the $(6,11,85)$-DGR in \cite{Shearer}.

\newpage

\begin{table}[h]
\caption{Computing $H(I,10)$ based on $(I-1, 10, H(I-1,10))$-DGR $(I=7,8,9,10).$}\label{table:algo}
\centering
\begin{tabular}{|c|c|c|c|c|}
\hline
$I$& Disjoint Golomb Rulers &$k$ & $b$ & $H(I,10)$\\
\hline
6& \uwave{12, 13, 18, 22, 35, 38, 46, 53, 65, 67;} &1& -2 & 70\\
&7, 9, 24, 31, 34, 50, 54, 62, 63, 68;&&&\\
&5, 8, 19, 21, 41, 42, 47, 51, 59, 66;&&&\\
&4, 6, 14, 20, 23, 43, 44, 48, 55, 70;&&&\\
&3, 10, 15, 25, 28, 29, 52, 58, 60, 69;&&&\\
&1, 2, 11, 16, 37, 39, 45, 57, 61, 64.&&&\\
\hline
7& \uwave{17, 19, 31, 38, 46, 49, 62, 66, 71, 72;} &3&1 &74 \\
&\textbf{\uwave{10, 11, 16, 20, 33, 36, 44, 51, 63, 65;}}&&&\\
&5, 6, 8, 12, 23, 35, 43, 48, 57, 67;&&&\\
&4, 14, 21, 26, 32, 34, 55, 58, 59, 74;&&&\\
&3, 7, 9, 25, 30, 37, 45, 56, 69, 70;&&&\\
&\uwave{2, 13, 22, 28, 29, 47, 50, 52, 60, 64;}&&&\\
&1, 15, 18, 24, 40, 42, 53, 61, 68, 73.&&&\\
\hline
8& \textbf{18, 20, 32, 39, 47, 50, 63, 67, 72, 73;} & 4 & 0 & 80 \\
&\textbf{11, 12, 17, 21, 34, 37, 45, 52, 64, 66;}&&&\\
&\uwave{8, 10, 19, 25, 26, 46, 49, 54, 68, 80;}&&&\\
&\uwave{6, 9, 16, 27, 35, 40, 60, 62, 76, 77;}&&&\\
&\uwave{5, 7, 15, 22, 28, 31, 56, 70, 74, 75;}&&&\\
&\uwave{4, 13, 36, 38, 42, 43, 55, 58, 69, 79;}&&&\\
&\textbf{3, 14, 23, 29, 30, 48, 51, 53, 61, 65;}&&&\\
&1, 2, 24, 33, 41, 44, 57, 59, 71, 78.&&&\\
\hline
9& \uwave{29, 34, 37, 48, 52, 61, 73, 83, 89, 90; }&5&0&90\\
&\uwave{18, 21, 23, 32, 39, 47, 59, 72, 78, 82;}&&&\\
&\textbf{8, 10, 19, 25, 26, 46, 49, 54, 68, 80;}&&&\\
&\textbf{6, 9, 16, 27, 35, 40, 60, 62, 76, 77;}&&&\\
&\textbf{5, 7, 15, 22, 28, 31, 56, 70, 74, 75;}&&&\\
&\textbf{4, 13, 36, 38, 42, 43, 55, 58, 69, 79;}&&&\\
&\uwave{3, 12, 17, 30, 33, 41, 45, 64, 84, 86;}&&&\\
&\uwave{2, 11, 14, 44, 50, 57, 65, 67, 81, 85;}&&&\\
&\uwave{1, 20, 24, 51, 53, 63, 66, 71, 87, 88.}&&&\\
\hline
10& 36, 46, 55, 60, 68, 80, 91, 95, 97, 98;& & &100\\
&31, 35, 40, 43, 54, 56, 76, 93, 94, 100;&&&\\
&\textbf{29, 34, 37, 48, 52, 61, 73, 83, 89, 90;}&&&\\
&\textbf{18, 21, 23, 32, 39, 47, 59, 72, 78, 82;}&&&\\
&8, 13, 16, 22, 26, 38, 49, 75, 77, 96;&&&\\
&5, 7, 10, 19, 25, 58, 62, 69, 79, 92;&&&\\
&4, 6, 9, 15, 27, 28, 42, 70, 74, 99;&&&\\
&\textbf{3, 12, 17, 30, 33, 41, 45, 64, 84, 86;}&&&\\
&\textbf{2, 11, 14, 44, 50, 57, 65, 67, 81, 85;}&&&\\
&\textbf{1, 20, 24, 51, 53, 63, 66, 71, 87, 88.}&&&\\
\hline
\end{tabular}

\end{table}

Exact values of  $H(I,J)$ and upper bounds on $H(I,J)$ for $7 \leq I \leq 13, 10 \leq J \leq 13$ are listed in Table \ref{table:UHIJ}, which are obtained without much difficulty. Upper bounds on $H(9,11)$, $H(10,12)$ and $H(12,13)$  are very close to the estimated exact values. Although longer computation may prompt more better upper bounds, we stopped because that getting better computation results is not our major concern in this paper. But note that, it seems not easy to improve these upper bounds without much computation. DGR achieving the upper bounds on $H(I, J)$ in Table \ref{table:UHIJ} for $I \leq J$, are shown in Table \uppercase\expandafter{\romannumeral4} in the Appendix.

\begin{center}

\begin{threeparttable} [H]

\footnotesize
\caption{Upper bounds on $H(I,J)$} \label{table:UHIJ}
 \begin{tabular}{p{0.5cm}|p{0.5cm}p{0.5cm}p{0.5cm}p{0.5cm}p{0.5cm}p{0.5cm}p{0.5cm}p{0.5cm}}
   \hline
   $J/I$ &  7&8&9 & 10 & 11 & 12 & 13\\
   \hline
   10 & \textbf{74}\tnote{*}  & \textbf{80} & \textbf{90} & \textbf{100} & \textbf{110} & \textbf{120} & \textbf{130}\\
   11 & \textbf{88}&  94&  100&  \textbf{110}&   \textbf{121} & \textbf{132} & \textbf{143}\\
   12 & \textbf{105} & 109 & 115 & 122&  \textbf{132} & \textbf{144} & \textbf{156}\\
   13 & \textbf{124} & 130 & 135&  141&  148 & 158 & \textbf{169} \\
    \hline

 \end{tabular}
 \begin{tablenotes}
  \item[*] Exact values of $H(I,J)$ are in bold.
\end{tablenotes}
\end{threeparttable}
\end{center}
\vspace{0.7cm}

\subsection{Bounds on $\tau(J)$}
Since a number of exact values of $H(I,J)$ have been determined for regular cases, only a little more effort should be made to improve the bounds on $\tau(J)$.  Observe that if $H(I_0,J)=I_0J$ has been determined, a $(2I_0, J, 2I_0J)$-DGR can be constructed by a union of an $(I_0, J, I_0J)$-DGR and its $(I_0J)$-step transformation. As a consequence, $H(I,J)$ need only to be determined for $I \in \{I_0+1, \cdots, 2I_0 -1\}$. For example, since it has been proved that $\tau(10) \in \{7,8,\cdots,20 \}$\cite{Klove}, $H(7,10)=74$ and $H(I,10)=10I$ where $8 \leq I \leq 13$, only $H(I,10)$ for $14 \leq I \leq 15$ need to be determined. Applying the method mentioned in the previous subsection, $H(14,10)=140$ and $H(15,10)=150$ are obtained without difficulty. Therefore $\tau(10) = 8$.
Bounds on $\tau(J)$ for $J \in \{11, 12, 13\}$ are improved by the same way, which are shown in Table \ref{table:tao}.

\begin{table} [h]
\caption{Bounds on $\tau(J)$ for $10 \leq J \leq 13$} \label{table:tao}
\centering
 \begin{tabular}{ccccc}

   \hline
   $J$ & 10 & 11 & 12 & 13\\
   \hline
   $\tau(J)$ & 8 & 9-10 & 9-11 & 10-13\\
    \hline

 \end{tabular}
\end{table}

By Table \ref{table:tao} and known values of $\tau(J)$, both Conjecture \ref{genefirstn2} and Conjecture \ref{HII2regular} hold for $J \leq 10$, and Conjecture \ref{genefirstn2}  holds for $J = 11$. It is still a pity, however, that values of $H(I, I+2)$ for $I \in \{9,10,11\}$ have not been determined yet.

\subsection{Computational Verification of Conjecture \ref{genefirstnb}}

We know that $H(4,5)=20$, that is, $(4,5,20)$-DGR is regular. By Conjecture \ref{genefirstnb}, for any 5-mark Golomb ruler $A$, there exists a $(5,5,25)$-DGR containing $A$, which has been confirmed by computer search. Similar computation for $H(5,5)$, $H(6,5)$, $H(5,6)$, $H(6,6)$ and $H(6,7)$ are also performed to confirm the conjecture. Note that there are more than 32 million 8-mark Golomb rulers from $\{1,2,\cdots,64\}$. It seems not easy to compute for the case of $H(7,8)$.

Moreover, the difference of the difficulty of search for DGR in different cases also suggests the correctness of the conjecture. Determining $H(I,J)$ where $I > J-2$ are much easier than determining $H(J-3,J)$ and $H(J-2,J)$. For example, we can prove that $H(13,13) = 169$ by computer search without much difficulty, but to determine $H(11,13) = 143$ is much more difficult, which is not achieved yet in this paper. For another example, $(I+1,12,12(I+1))$-DGRs where $11 < I < 21$ can be easily discovered by the proposed method , even when specifying $k=I-2$.

\section{Conclusions and Remarks} \label{conclu}
The Golomb Ruler Problem consists in finding a set of distinct non-negative integers such that all differences between pairs of integers are distinct, while minimizing the largest difference. Finding disjoint Golomb rulers is an interesting generalization of the Golomb ruler problem.  We generalize the problem to arbitrary $n$ positive integers and conjecture that there exist $I$ disjoint Golomb rulers, each being a $J$-subset of any set $A$ of positive integers such that $|A| = H(I,J)$. We have proved that the conjecture can deduce some more interesting conjectures, based on which upper bounds on $H(I,J)$ for  $7 \leq I \leq 13$ and $10 \leq J \leq 13$ are obtained by computer search. Moveover, a conjecture about Golomb ruler is proposed, which is stronger than the old one mentioned first by Erd\H{o}s. Theoretical proofs and computational verification of these conjecture are our future tasks.

\newpage
\appendix[Disjoint Golomb Rulers Achieving Upper Bounds on $H(I,J)$]
 \vspace{0.3cm}
\begin{center}
\footnotesize TABLE \uppercase\expandafter{\romannumeral4}

 S\scriptsize ETS OF DISJOINT GOLOMB RULERS
\end{center}

\begin{center}
\footnotesize
 \tablefirsthead{\hline$J$ &  $I$ & \multicolumn{13}{l|} {Disjoint Golomb Rulers}\\
   &&&&&&&&&&&&&&\\\hline }
  \tablelasttail{\hline}

\label{table:sdgr}

\tablehead{\hline$J$ &  $I$ & \multicolumn{13}{l|} {Disjoint Golomb Rulers}\\
   &&&&&&&&&&&&&&\\\hline }
\tabletail{\hline\multicolumn{15}{ r }{\textit{(continued on the next page)}}\\}
 \label{table:sdgr}
 \begin{supertabular}[r]{|p{0.3cm}p{0.35cm}p{0.35cm}p{0.35cm}p{0.35cm}p{0.35cm}p{0.35cm}p{0.35cm}p{0.35cm}p{0.35cm}p{0.35cm}p{0.35cm}p{0.35cm}p{0.35cm}p{0.45cm}<{\centering}|}
   \hline

   10 & 7&17&19&31&38&46&49&62&66&71&72&&&\\
     &&10&11&16&20&33&36&44&51&63&65&&&\\
    &&5&6&8&12&23&35&43&48&57&67&&&\\
    &&4&14&21&26&32&34&55&58&59&74&&&\\
    &&3&7&9&25&30&37&45&56&69&70&&&\\
    &&2&13&22&28&29&47&50&52&60&64&&&\\
    &&1&15&18&24&40&42&53&61&68&73&&&\\

    \hline
10& 8& 8& 10& 19& 25& 26& 46& 49& 54& 68& 80&&&\\
&&6& 9& 16& 27& 35& 40& 60& 62& 76& 77&&&\\
&&5& 7& 15& 22& 28& 31& 56& 70& 74& 75&&&\\
&&4& 13& 36& 38& 42& 43& 55& 58& 69& 79&&&\\
&&1& 2& 24& 33& 41& 44& 57& 59& 71& 78&&&\\
&&18& 20& 32& 39& 47& 50& 63& 67& 72& 73&&&\\
&&11& 12& 17& 21& 34& 37& 45& 52& 64& 66&&&\\
&&3& 14& 23& 29& 30& 48& 51& 53& 61& 65&&&\\

\hline
10&9& 29&34&37&48&52&61&73&83&89&90&&&\\
&&18&21&23&32&39&47&59&72&78&82&&&\\
&&3&12&17&30&33&41&45&64&84&86&&&\\
&&2&11&14&44&50&57&65&67&81&85&&&\\
&&1&20&24&51&53&63&66&71&87&88&&&\\
&&8&10&19&25&26&46&49&54&68&80&&&\\
&&6&9&16&27&35&40&60&62&76&77&&&\\
&&5&7&15&22&28&31&56&70&74&75&&&\\
&&4&13&36&38&42&43&55&58&69&79&&&\\

\hline
10&10& 36&46&55&60&68&80&91&95&97&98&&&\\
&&31&35&40&43&54&56&76&93&94&100&&&\\
&&8&13&16&22&26&38&49&75&77&96&&&\\
&&5&7&10&19&25&58&62&69&79&92&&&\\
&&4&6&9&15&27&28&42&70&74&99&&&\\
&&29&34&37&48&52&61&73&83&89&90&&&\\
&&18&21&23&32&39&47&59&72&78&82&&&\\
&&3&12&17&30&33&41&45&64&84&86&&&\\
&&2&11&14&44&50&57&65&67&81&85&&&\\
&&1&20&24&51&53&63&66&71&87&88&&&\\

\hline
11&7&13& 14& 17& 22& 36& 43& 54& 56& 71& 81& 87&&\\
&&11& 15& 18& 34& 35& 40& 49& 67& 75& 77& 88&&\\
&&7& 10& 21& 23& 27& 48& 55& 60& 70& 78& 79&&\\
&&4& 16& 19& 20& 45& 52& 58& 63& 72& 80& 82&&\\
&&3& 8& 24& 30& 33& 37& 47& 65& 73& 84& 85&&\\
&&2& 9& 25& 26& 28& 39& 59& 64& 68& 74& 86&&\\
&&1& 5& 6& 12& 32& 41& 51& 53& 66& 69& 83&&\\

\hline
11&8& 9& 14& 23& 26& 39& 57& 59& 63& 78& 85& 86&&\\
&&7& 11& 32& 40& 41& 52& 55& 68& 87& 92& 94&&\\
&&4& 5& 27& 29& 34& 46& 62& 66& 72& 80& 93&&\\
&&3& 10& 25& 33& 35& 44& 70& 71& 75& 88& 91&&\\
&&1& 2& 8& 12& 28& 30& 43& 64& 67& 76& 81&&\\
&&15& 16& 19& 24& 38& 45& 56& 58& 73& 83& 89&&\\
&&13& 17& 20& 36& 37& 42& 51& 69& 77& 79& 90&&\\
&&6& 18& 21& 22& 47& 54& 60& 65& 74& 82& 84&&\\

\hline
11&9&13&26&36&42&43&61&63&85&89&94&97&&\\
&&10&17&29&30&32&40&58&75&91&96&100&&\\
&&4&7&16&21&51&55&57&70&77&78&88&&\\
&&2&5&12&28&33&45&47&60&69&98&99&&\\
&&1&3&6&41&50&54&65&66&73&87&93&&\\
&&9& 18& 34& 35& 37& 48& 68& 72& 80& 90& 95&&\\
&&15& 19& 22& 38& 39& 44& 53& 71& 79& 81& 92&&\\
&&11& 14& 25& 27& 31& 52& 59& 64& 74& 82& 83&&\\
&&8& 20& 23& 24& 49& 56& 62& 67& 76& 84& 86&&\\

\hline
11&10&11& 19& 31& 37& 52& 53& 88& 98& 102& 105& 107&&\\
&&1& 6& 7& 9& 35& 48& 62& 66& 73& 83& 106&&\\
&&2& 5& 27& 38& 40& 50& 54& 69& 101& 108& 109&&\\
&&3& 4& 8& 28& 30& 46& 58& 87& 97& 104& 110&&\\
&&10& 15& 22& 32& 36& 55& 68& 70& 71& 79& 99&&\\
&&12& 16& 43& 57& 64& 65& 75& 77& 94& 100& 103&&\\
&&18& 23& 24& 34& 41& 49& 61& 63& 82& 91& 95&&\\
&&13& 17& 26& 45& 47& 59& 67& 74& 84& 85& 90&&\\
&&20& 21& 29& 39& 44& 51& 72& 76& 78& 89& 92&&\\
&&14& 25& 33& 42& 56& 60& 80& 81& 86& 93& 96&&\\

\hline
11&11&16& 30& 38& 47& 49& 50& 74& 79& 89& 95& 102&&\\
&&17& 21& 39& 44& 55& 58& 70& 83& 90& 91& 100&&\\
&&20& 26& 36& 41& 65& 66& 68& 77& 85& 99& 103&&\\
&&28& 29& 33& 40& 43& 60& 76& 78& 84& 97& 106&&\\
&&19& 22& 23& 42& 48& 56& 80& 87& 96& 98& 108&&\\
&&14& 15& 25& 45& 69& 72& 81& 86& 88& 94& 109&&\\
&&18& 24& 31& 32& 35& 51& 61& 73& 82& 105& 107&&\\
&&4& 9& 12& 37& 53& 54& 64& 93& 111& 115& 117&&\\
&&3& 6& 10& 46& 57& 62& 71& 92& 110& 112& 120&&\\
&&1& 7& 8& 27& 52& 63& 67& 104& 113& 116& 121&&\\
&&2& 5& 11& 13& 34& 59& 75& 101& 114& 118& 119&&\\

\hline
12&7& 6& 13& 35& 37& 40& 46& 56& 76& 84& 88& 101& 102&\\
&&1& 2& 20& 24& 44& 52& 59& 69& 85& 90& 96& 99&\\
&&3& 8& 11& 15& 25& 31& 60& 61& 79& 92& 94& 103&\\
&&4& 7& 9& 27& 41& 53& 70& 74& 80& 81& 89& 105&\\
&&5& 16& 17& 19& 26& 50& 55& 63& 78& 82& 98& 104&\\
&&10& 14& 28& 36& 45& 47& 48& 72& 77& 87& 93& 100&\\
&&12& 21& 22& 29& 42& 54& 57& 68& 73& 91& 95& 97&\\

\hline
12&8&10& 12& 16& 34& 39& 50& 53& 65& 78& 85& 86& 95&\\
&&8& 9& 24& 29& 32& 57& 59& 66& 70& 76& 88& 102&\\
&&5& 11& 14& 19& 41& 42& 62& 80& 87& 91& 104& 106&\\
&&3& 7& 21& 26& 36& 38& 47& 60& 63& 101& 108& 109&\\
&&2& 4& 25& 31& 44& 49& 81& 82& 90& 93& 97& 107&\\
&&1& 15& 33& 35& 48& 72& 73& 77& 84& 94& 100& 103&\\
&&6& 17& 18& 20& 27& 51& 56& 64& 79& 83& 99& 105&\\
&&13& 22& 23& 30& 43& 55& 58& 69& 74& 92& 96& 98&\\

\hline
12&9&17& 26& 27& 34& 47& 59& 62& 73& 78& 96& 100& 102&\\
&&14& 16& 20& 38& 43& 54& 57& 69& 82& 89& 90& 99&\\
&&12& 13& 28& 33& 36& 61& 63& 70& 74& 80& 92& 106&\\
&&10& 21& 22& 24& 31& 55& 60& 68& 83& 87& 103& 109&\\
&&5& 19& 37& 39& 52& 76& 77& 81& 88& 98& 104& 107&\\
&&6& 7& 15& 30& 35& 42& 46& 48& 67& 97& 111& 114&\\
&&4& 8& 23& 29& 40& 53& 56& 79& 84& 91& 93& 113&\\
&&2& 9& 41& 45& 51& 64& 85& 86& 101& 110& 112& 115&\\
&&1& 3& 18& 44& 50& 66& 71& 75& 94& 95& 105& 108&\\

\hline
12&10&5& 6& 29& 36& 61& 65& 70& 80& 82& 108& 119& 122&\\
&&3& 4& 8& 10& 27& 37& 49& 63& 81& 84& 112& 121&\\
&&2& 7& 9& 15& 57& 60& 76& 77& 86& 98& 109& 123&\\
&&1& 24& 26& 42& 46& 54& 55& 88& 93& 103& 114& 120&\\
&&22& 31& 32& 39& 52& 64& 67& 78& 83& 101& 105& 107&\\
&&19& 21& 25& 43& 48& 59& 62& 74& 87& 94& 95& 104&\\
&&17& 18& 33& 38& 41& 66& 68& 75& 79& 85& 97& 111&\\
&&14& 20& 23& 28& 50& 51& 71& 89& 96& 100& 113& 115&\\
&&12& 16& 30& 35& 45& 47& 56& 69& 72& 110& 117& 118&\\
&&11& 13& 34& 40& 53& 58& 90& 91& 99& 102& 106& 116&\\
\hline
12&11&14&16&20&38&43&54&57&69&82&89&90&99&\\
&&12&13&28&33&36&61&63&70&74&80&92&106&\\
&&9&15&18&23&45&46&66&84&91&95&108&110&\\
&&7&11&25&30&40&42&51&64&67&105&112&113&\\
&&6&8&29&35&48&53&85&86&94&97&101&111&\\
&&5&19&37&39&52&76&77&81&88&98&104&107&\\
&&27&34&49&62&73&87&103&120&124&129&130&132&\\
&&17&21&41&50&60&68&100&102&116&117&123&128&\\
&&4&24&32&47&56&58&72&109&114&121&127&131&\\
&&2&3&26&31&44&65&71&79&96&115&122&126&\\
&&1&10&22&55&59&75&78&83&93&118&119&125&\\
\hline
12&12&33& 37& 40& 63& 72& 77& 88& 101& 122& 134& 142& 144&\\
&&1& 4& 5& 23& 32& 61& 75& 86& 109& 133& 139& 141&\\
&&2& 3& 9& 13& 25& 55& 57& 81& 102& 115& 140& 143&\\
&&7& 15& 22& 24& 34& 48& 87& 103& 107& 132& 137& 138&\\
&&27& 31& 45& 53& 62& 64& 65& 89& 94& 104& 110& 117&\\
&&29& 38& 39& 46& 59& 71& 74& 85& 90& 108& 112& 114&\\
&&28& 35& 41& 56& 58& 68& 82& 100& 111& 116& 119& 120&\\
&&16& 17& 21& 43& 52& 54& 66& 98& 105& 113& 123& 126&\\
&&14& 19& 36& 44& 50& 76& 79& 91& 95& 118& 128& 129&\\
&&10& 11& 42& 49& 51& 67& 70& 78& 84& 121& 131& 136&\\
&&18& 26& 30& 47& 69& 83& 92& 93& 99& 125& 127& 130&\\
&&6& 8& 12& 20& 60& 73& 80& 96& 97& 106& 124& 135&\\
\hline
13&7&12& 13& 15& 35& 44& 54& 59& 70& 97& 104& 110& 118& 122\\
&&7& 8& 20& 24& 27& 51& 53& 62& 87& 101& 109& 119& 124\\
&&5& 11& 16& 29& 30& 52& 69& 78& 81& 85& 113& 115& 123\\
&&4& 19& 25& 33& 45& 55& 58& 89& 93& 98& 100& 116& 117\\
&&3& 6& 14& 41& 43& 50& 65& 75& 91& 95& 96& 108& 114\\
&&2& 9& 23& 26& 28& 34& 64& 68& 84& 99& 111& 112& 121\\
&&1& 10& 17& 22& 37& 61& 72& 74& 80& 102& 103& 106& 120\\
\hline
13&8&15& 16& 32& 34& 39& 43& 74& 77& 87& 99& 107& 113& 128\\
&&1& 11& 24& 28& 29& 36& 70& 73& 94& 103& 109& 123& 125\\
&&2& 3& 6& 27& 40& 46& 69& 76& 91& 96& 108& 122& 124\\
&&5& 13& 30& 41& 62& 67& 81& 85& 105& 112& 114& 115& 127\\
&&7& 19& 21& 22& 26& 51& 59& 72& 90& 100& 106& 117& 126\\
&&8& 9& 20& 23& 25& 33& 54& 60& 80& 98& 102& 121& 130\\
&&4& 10& 18& 31& 35& 42& 61& 64& 79& 84& 119& 120& 129\\
&&12& 14& 17& 37& 49& 55& 71& 82& 97& 101& 110& 111& 118\\
\hline
13&9&21& 26& 30& 38& 48& 59& 62& 85& 101& 116& 129& 135& 136\\
&&1& 4& 5& 29& 41& 47& 76& 91& 96& 98& 107& 117& 130\\
&&2& 9& 32& 45& 46& 49& 67& 95& 100& 115& 124& 126& 134\\
&&3& 8& 12& 20& 34& 40& 78& 79& 81& 97& 108& 121& 131\\
&&6& 13& 27& 37& 53& 55& 56& 88& 110& 118& 122& 127& 133\\
&&18& 23& 24& 43& 54& 58& 61& 70& 87& 109& 111& 119& 132\\
&&10& 15& 17& 25& 31& 64& 65& 82& 93& 102& 106& 125& 128\\
&&7& 11& 22& 28& 35& 36& 69& 72& 74& 92& 104& 114& 123\\
&&14& 16& 19& 39& 51& 57& 73& 84& 99& 103& 112& 113& 120\\
\hline
13&10&9& 15& 35& 42& 43& 57& 73& 82& 86& 118& 136& 139& 141\\
&&1& 4& 5& 28& 40& 69& 78& 84& 89& 103& 121& 129& 131\\
&&2& 21& 37& 45& 54& 68& 74& 75& 79& 123& 125& 135& 138\\
&&3& 18& 22& 31& 39& 53& 77& 95& 97& 100& 107& 134& 140\\
&&7& 8& 23& 27& 34& 67& 85& 91& 99& 108& 120& 130& 133\\
&&6& 14& 32& 41& 47& 61& 66& 83& 104& 111& 114& 115& 127\\
&&11& 13& 16& 36& 48& 49& 55& 65& 76& 106& 110& 124& 132\\
&&19& 25& 30& 38& 52& 56& 80& 81& 96& 116& 119& 126& 128\\
&&20& 24& 26& 29& 51& 64& 72& 87& 88& 98& 105& 117& 137\\
&&10& 12& 17& 33& 44& 50& 58& 70& 94& 109& 112& 113& 122\\
\hline
13&11&30& 34& 48& 59& 60& 83& 99& 105& 126& 133& 141& 143& 146\\
&&6& 31& 40& 43& 57& 70& 72& 78& 88& 124& 128& 147& 148\\
&&5& 18& 24& 26& 33& 55& 66& 71& 96& 110& 122& 142& 145\\
&&3& 4& 7& 25& 56& 65& 85& 93& 104& 119& 129& 131& 136\\
&&2& 8& 22& 38& 46& 77& 80& 87& 89& 102& 106& 134& 139\\
&&1& 10& 11& 37& 41& 62& 76& 82& 84& 95& 100& 132& 144\\
&&9& 17& 35& 44& 50& 64& 69& 86& 107& 114& 117& 118& 130\\
&&14& 16& 19& 39& 51& 52& 58& 68& 79& 109& 113& 127& 135\\
&&23& 27& 29& 32& 54& 67& 75& 90& 91& 101& 108& 120& 140\\
&&12& 21& 42& 45& 49& 92& 98& 103& 111& 121& 123& 137& 138\\
&&13& 15& 20& 36& 47& 53& 61& 73& 97& 112& 115& 116& 125\\
\hline
13&12&24& 38& 62& 74& 96& 104& 111& 129& 139& 150& 152& 155& 156\\
&&1& 11& 12& 26& 56& 78& 94& 98& 106& 115& 141& 147& 154\\
&&2& 3& 16& 31& 47& 51& 57& 81& 84& 89& 140& 151& 158\\
&&4& 6& 18& 41& 59& 80& 83& 93& 102& 108& 109& 113& 153\\
&&5& 13& 28& 48& 49& 55& 67& 77& 101& 132& 146& 148& 157\\
&&22& 29& 30& 39& 43& 58& 69& 85& 91& 103& 123& 126& 128\\
&&23& 27& 32& 33& 50& 52& 64& 90& 97& 112& 125& 133& 136\\
&&20& 21& 35& 42& 44& 61& 72& 88& 92& 117& 127& 130& 135\\
&&19& 36& 37& 45& 60& 65& 87& 99& 118& 120& 124& 131& 134\\
&&25& 40& 46& 54& 66& 76& 79& 110& 114& 119& 121& 137& 138\\
&&7& 10& 14& 15& 53& 71& 82& 95& 107& 116& 122& 142& 144\\
&&8& 9& 17& 34& 63& 70& 73& 86& 100& 105& 143& 145& 149\\
\hline
13&13&3& 6& 19& 20& 59& 70& 85& 97& 106& 116& 158& 160& 165\\
&&30& 37& 38& 47& 51& 66& 77& 93& 99& 111& 131& 134& 136\\
&&31& 35& 40& 41& 58& 60& 72& 98& 105& 120& 133& 141& 144\\
&&28& 29& 43& 50& 52& 69& 80& 96& 100& 125& 135& 138& 143\\
&&27& 44& 45& 53& 68& 73& 95& 107& 126& 128& 132& 139& 142\\
&&33& 48& 54& 62& 74& 84& 87& 118& 122& 127& 129& 145& 146\\
&&15& 18& 22& 23& 61& 79& 90& 103& 115& 124& 130& 150& 152\\
&&16& 17& 25& 42& 71& 78& 81& 94& 108& 113& 151& 153& 157\\
&&9& 11& 14& 46& 55& 65& 76& 91& 114& 148& 154& 161& 162\\
&&12& 21& 26& 56& 63& 64& 67& 88& 121& 137& 149& 166& 168\\
&&1& 7& 24& 34& 39& 75& 83& 101& 104& 117& 147& 156& 167\\
&&2& 4& 13& 36& 49& 82& 89& 92& 109& 110& 140& 159& 164\\
&&5& 8& 10& 32& 57& 86& 102& 112& 119& 123& 155& 163& 169\\
\hline
\end{supertabular}
\end{center}



\ifCLASSOPTIONcaptionsoff
  \newpage
\fi



%

\bibliographystyle{IEEEtran}
\bibliography{IEEEabrv,dgr}

%

\begin{IEEEbiographynophoto}{Baoxin Xiu}
received the B.Sc. degree in applied mathematics in 2000, and the Ph.D. degree in management science and engineering in 2006, both from   National University of Defense Technology (NUDT), Changsha, China. He is currently an associate professor in Information Systems Engineering Laboratory at NUDT. His research interests include algorithm design, granular computing and complex network.
\end{IEEEbiographynophoto}
\begin{IEEEbiographynophoto}{Changjun Fan}
received the B.Sc. degree in command information system from National University of Defense Technology, Changsha, China, in 2013. He is currently a M.Sc. candidate in the College of Information System and Management at National University of Defense Technology, Changsha, China. His research interests include complex network and data mining.
\end{IEEEbiographynophoto}
\begin{IEEEbiographynophoto}{Meilian Liang}
received her B.Sc. degree and M.Sc. degree in computer science from Guangxi University in 2002
and 2005, respectively. She is currently an associate professor of Computer Science at Guangxi University, Nanning. Her
research interests include data mining and algorithm design.
\end{IEEEbiographynophoto}




\end{document}